\documentclass[12pt,a4paper]{article}
\usepackage[utf8x]{inputenc}
\usepackage[T1]{fontenc}

\usepackage{amsmath}
\usepackage{textcomp, gensymb}
\usepackage{caption}
\usepackage{braket}
\usepackage{dsfont}
\usepackage{amsfonts}
\usepackage[english]{babel}
\usepackage{setspace}

\usepackage{yhmath} 

\usepackage[style=numeric, backend=biber, doi=true, url=true, maxbibnames=99,minbibnames=99]{biblatex}
\usepackage{hyperref}
\usepackage{graphicx}
\usepackage{enumitem} 

\usepackage[a4paper, lmargin=3cm, rmargin=3cm, tmargin=2cm, bmargin=2cm]{geometry} 
\usepackage[all]{nowidow} 

\title{Using self-testing for semi-device-independent certification of qualitative properties: an analysis of asymmetry certification}
\author{Gergő Veres$^1$ and Gábor Drótos$^{2,3}$}
\date{%
    \textit{%
        $^1$University of Debrecen, Debrecen, Hungary\\%
        $^2$Insituto de Física Interdisciplinar y Sistemas Complejos (IFISC; CSIC-UIB), Palma de Mallorca, Spain\\%
        $^3$Insitute for Nuclear Research (ATOMKI), Debrecen, Hungary\\[2ex]%
    }%
    \today
}

\begin{document} 
\maketitle

\begin{abstract}
    Certifying a single qualitative property of a configuration of qubits in a prepare-and-measure scenario is a relatively unexplored direction in semi-device-independent (semi-DI) certification. Assuming an upper bound on the dimension of the quantum system involved, we analyse here properties of such a task. Similarly to Ref.~\cite{TaoutiouiEtAl2025}, we use a semi-DI self-testing construction, but a different one, to introduce a linear witness to certify the asymmetry of a three-qubit configuration (with one further qubit added for technical reasons). Certification is achieved by constructing a self-test for an asymmetric configuration and determining (numerically) the maximum witness value accessible by any symmetric configuration (with the three Bloch vectors under consideration describing an isosceles triangle). We will first ask whether the gap between these two witness values can serve as a quantifier of asymmetry. More generally, we analyse the landscape of the witness gap over the parameter space of the asymmetric configuration with various parameter choices of the self-testing construction. We numerically find that the landscape is smooth, has a simple shape and is structurally robust to alterations of the parameters. However, since the gap depends on the size of the triangle defined by the qubit Bloch vectors and on the construction chosen for self-testing, this gap is not a reasonable quantifier of asymmetry alone. Notwithstanding, certifying the asymmetric nature (or other qualitative property) of a three-qubit configuration according to our approach is presumably easier at larger gaps, exploitation of which may be facilitated by the favourable properties of the landscape.
\end{abstract}

\textit{keywords: semi-device-independent certification, prepare-and-measure scenario, mirror symmetry}

\section{Introduction}

Besides quantum technological applications, certification protocols related to quantum information play an important role in research addressing the basic physical laws of nature. Certification would ideally imply avoiding the use of any assumptions, but this is not possible in the physical world. For instance, a usual assumption in certifying the desired working or some interesting property of a real-world setup is that quantum mechanics describes the physics of the system appropriately. As an example, experimental evidence was recently provided in Refs.~\cite{LiEtAl2022,ChenEtAl2022} in a multipartite Bell-like setup that a quantum mechanical description must admit superpositions in complex systems that require properties of complex numbers; the setup and the certification scheme was proposed in Ref.~\cite{RenouEtAl2021} a year earlier in a theoretical work.

The certification task realised in the referred studies in fact aligns into the so-called device-independent (DI) framework \cite{AcinEtAl2007,Scarani2012}. This framework is sometimes described as lacking any assumptions, which in fact refers to the lack of assumptions about the state preparations and measurements involved. Device-independent protocols rely on entanglement, which can be avoided by making some simple assumption about the prepared states, in which case we speak about a semi-device-independent approach and which is most often turned to in a prepare-and-measure (PM) scenario \cite{PawlowskiBrunner2011,VanHimbeeckEtAl2017}.

In some sense, the most stringent certification task formulated in a DI or semi-DI framework is called self-testing; it means certifying that an unknown configuration of prepared states and/or measurement settings is equivalent to a desired one up to an isometry (unitary or anti-unitary transformation if dimensionality is preserved) \cite{MayersYao2004,SupicBowles2020}. Since perfect equivalence is impossible to reach in the real world, certification of ``epsilon-closeness'', known as robust self-testing, is also often considered \cite{McKagueEtAl2012}. Self-testing and robust self-testing are subject to intensive research partially because many quantum information protocols envisioning real-world applications rely on specific configurations (see e.g. Ref.~\cite{CaoEtAl2022}).

However, certifying much less about the setup, such as some single, in some sense qualitative global property, may also be of interest, as the case of complex superpositions illustrates. This approach has been in intuitive use since Ref.~\cite{TavakoliEtAl2018} at least and was explicitly discussed in a recent work of ours \cite{DrotosEtAl2024}. The usual idea is to choose some target configuration that possesses the desired property, construct a so-called witness functional on the space of (conditional) outcome probabilities such that its maximum value self-tests the target, then to determine the maximum of the witness attainable by configurations not possessing the desired property. If evaluating the witness for some unknown configuration results in exceeding the latter maximum, it is then a certification of the desired property.

In this spirit, certifying asymmetry (lack of geometric symmetry) in an unknown configuration of three qubits has been addressed in Ref.~\cite{TaoutiouiEtAl2025}. Besides illustrating the feasibility of the task, the gap between the ideal (self-testing) witness value and the maximum attainable by symmetric configurations (the `mirror-symmetric bound') has been proposed both to be a quantifier of asymmetry of the target and to characterise how "easy" it is to perform the certification, i.e., how much it is tolerant to imperfections. Our present paper aims at a deeper analysis of the issue, refining and challenging, to some extent, the mentioned ideas.

In particular, we follow the very same methodology as in Ref.~\cite{TaoutiouiEtAl2025} for most of our analysis but use a different construction for self-testing the target. We carefully investigate the landscape of the witness gap over different target configurations and conclude that this landscape is qualitatively similar to that in Ref.~\cite{TaoutiouiEtAl2025} but is quantitatively different. Perhaps the most conspicuous difference is the different location of the maximum gap within the landscape, meaning that the ``most asymmetric'' configuration depends on which construction is chosen for the self-test. While this clearly implies that the witness gap is not a universal quantifier of asymmetry, the issue of tolerance to imperfections necessitates more involved considerations. Note that tolerance to imperfections is related to but is not the same problem as which configurations may be certified to be asymmetric by the help of a given target.

The paper is organised as follows. After summarizing general background and a particular self-testing construction in Section~\ref{sec:pre}, we describe our way of analysing the issue in Section~\ref{sec:meth}. Findings in Section~\ref{sec:find} are presented first in a basic setup, then we alter a parameter and, finally, also alter a methodological choice. Our findings are discussed in comparison with Ref.~\cite{TaoutiouiEtAl2025} and corresponding conclusions are drawn in Section~\ref{sec:disc_concl}. Specifying and discussing some technical details are relegated to appendices.

\section{Preliminaries}\label{sec:pre}

\begin{figure}[!h]
        \centering
         \includegraphics[width=0.7\linewidth]{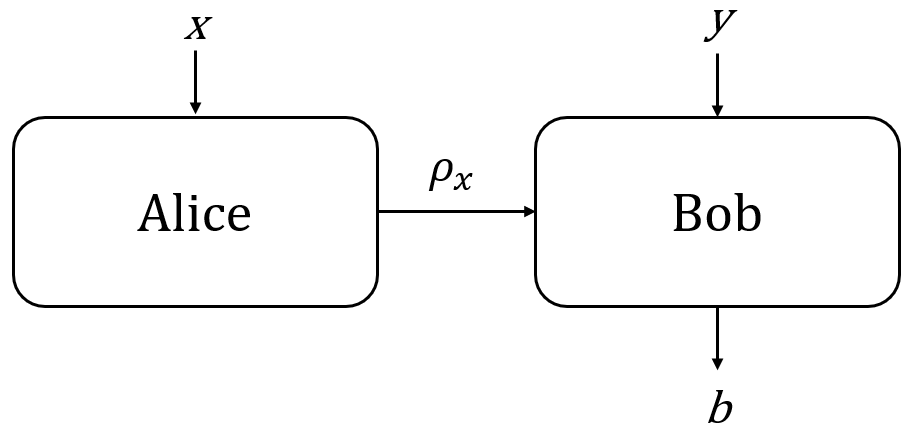}
		\caption{Schematic of the prepare-and-measure scenario, with the labels $x$, $y$ and $b$ as well as the transmitted state $\rho_x$ marked.}
        \label{fig:pm}
    \end{figure}

To avoid trust in preparations or measurement settings, semi-DI certification solely relies on output statistics arising from uncharacterised devices performing measurements, besides some physical assumption in addition (in our case, an upper bound on the relevant system dimension). Semi-DI certification is most often used in the traditional prepare-and-measure scenario, where Alice prepares quantum states $\rho_x$, upon receiving an input $x$, for measurements to be performed on them by Bob. The measurement setting chosen by Bob depends on another input $y$, so that Bob's measurements are described by the two-index family $(M_{b|y})_{b,y}$ of POVM elements, where $b$ is a label associated with the measurement outcome. We restrict ourselves to two-outcome measurements, i.e., $b\in\{1,2\}$. See Fig.~\ref{fig:pm} for an illustration of the scenario.

Self-testing practically means that we can tell whether the targeted states were prepared and the targeted measurements were carried out.
Self-testing, among other certification tasks, can be performed by defining a linear witness $W$ using the probabilities $P(b|xy) = Tr(\rho_x M_{b|y})$ that describe the measurement outcomes. Remember that these probabilities are invariant under unitary and anti-unitary transformations of the whole configuration, so that only equivalence to the target, up to such a transformation, may be certified.
    \begin{equation}
      W = \sum_{x=1}^{M_n} \sum_{y=1}^{M_m} \sum_{b=0}^{1}w_{xy}^{(b)}P(b|xy).
    \end{equation}
    
We intend to self-test some given preparations and measurements by the maximum witness value. For such a purpose, we can always choose $w_{xy}\equiv w_{xy}^{(0)} = - w_{xy}^{(1)}$ (these are the elements of what is called the witness matrix), and knowing that $P(0|xy)+P(1|xy)=1$, we have
    \begin{equation}
       W=\sum_{x=1}^{M_n} \sum_{y=1}^{M_m}w_{xy}[P(0|xy)-P(1|xy)].
    \end{equation}

For the purposes of our analyses and before specifying the witness matrix, we express the witness in terms of Bloch vectors entering the setup. The most general forms of a qubit state and a two-outcome measurement are
    \begin{align}
        \rho = \frac{\mathds{1}+\Vec{n} \cdot \Vec{\sigma}}{2}, \hspace{1cm} 
        M_b = [1+(-1)^b\mu]\frac{\mathds{1}+(-1)^b\Vec{m}^{(b)}\cdot\Vec{\sigma}}{2}
    \end{align}
where $\Vec{n}$ and $\Vec{m}$ are the Bloch vectors of preparations and measurements, respectively, and $\mu\in[-1,1]$ such that $(1+\mu)\vec{m}^{(0)} = (1-\mu)\vec{m}^{(1)}$. The self-testing construction will ensure that the maximum witness value cannot be attained with degenerate measurements, so that we take $\mu = 0$ for all of the measurements:
    \begin{equation}
        M_{b|y} = \frac{\mathds{1}+(-1)^b\Vec{m}\cdot\Vec{\sigma}}{2}.
    \end{equation}
Using all of the above equations we obtain
    \begin{equation}
        W = \sum_{x=1}^{M_n} \sum_{y=1}^{M_n}w_{xy} \Vec{n}_x \cdot \Vec{m}_y.
    \end{equation}

We also point out an important property. For a given $\Vec{m_y}$ the optimal choice for $\Vec{n_x}$ is
\begin{equation}
    \Vec{n_x} = \frac{\sum_{y=1}^{M_n} w_{xy} \Vec{m_y}}{\left|\sum_{y=1}^{M_n} w_{xy} \Vec{m_y}\right|} = \frac{\Vec{u_x}}{\left|\Vec{u_x}\right|},
\end{equation}
where $\Vec{u_x} \equiv \sum_{y=1}^{M_m}w_{xy}  \Vec{m_y}$, and if the denominator is not zero. This condition is true for every $x$ if $\Vec{m_y}$ is optimal and if every row of $w_{xy}$ contains a non-zero element. If these conditions are fulfilled, then $W$ is maximal regarding the preparations and can be computed in the following form:
\begin{equation} \label{eq:Qv}
    Q_m = \sum_{x=1}^{M_n} \left|\Vec{u_x}\right|.
\end{equation}
The last equation is also true if $\Vec{m_y}$ is such that $\left|\Vec{u_x}\right|=0$ for some $x$. In this instance any $\Vec{n_x}$ is optimal for a given $x$. This case is insignificant for us, because with the following construction we always determine the optimal measurements throughout our analyses, for which $\left|\Vec{u_x}\right|\neq 0$ (see Appendix B in Ref.~\cite{DrotosEtAl2024b}).

Now we choose a construction for self-testing that requires a low (perhaps minimal) number of preparations and measurements \cite{DrotosEtAl2024}. The number of measurements, $M_m$, must be equal to the dimension of the space spanned by the preparation Bloch vectors $\Vec{n}_x$. To ensure that the optimal witness value cannot be attained by degenerate measurements, $\sum_xw_{xy}=0$ must be fulfilled for all $y$. With this condition we can parametrize $w_{xy}$ in the following way:
    \begin{equation}
        w_{xy} = r_x\mu_{xy}
    \end{equation}
where $\sum_x\mu_{xy}=0$. To self-test a configuration of $M_n$ qubits, some positive numbers $r_x$ are chosen, then $\mu_{xy}$ must be chosen to be the components of $\Vec{n}_x$ in the coordinate system where the matrix $\sum_x r_x \Vec{n}_x \otimes \Vec{n}_x$ is diagonal. Choosing numbers $r_x$ such that $\sum_x r_x \Vec{n}_x = 0$ will ensure that degenerate measurements cannot be optimal. If this is not possible, then it is necessary but also enough to add one extra preparation in the form of
    \begin{equation} \label{eq:extraprep}
        -\sum_xr_x\Vec{n}_x.
    \end{equation}
The optimum will be unique if the condition $M_n \ge \frac{M_m(M_m-1)}{2}+1$ is fulfilled, which may also require adding an extra preparation.

A final remark to be made is about defining mirror symmetry. While the idea is related to simple geometry, there are two possible choices for a configuration of Bloch vectors: whether the origin is required to be in the symmetry plane or not. When one is concerned by properties of the quantum mechanical system, the former choice is desirable; at the same time, from a purely geometrical point of view, it is reasonable not to make this restriction. Ref.~\cite{TaoutiouiEtAl2025} makes the latter choice, but also proves in its Appendix~B that the optimal mirror symmetric configuration always consists of pure states with the particular self-testing construction used there, in which case the origin is in the symmetry plane. In our present paper, motivated by physics, we prefer to make the restriction of requiring the origin to be in the symmetry plane, but the special property just mentioned about Ref.~\cite{TaoutiouiEtAl2025} makes all of our results presented herein directly comparable with those of Ref.~\cite{TaoutiouiEtAl2025}.

\section{Methodology}\label{sec:meth}

For our study as outlined in the Introduction, we need to parametrize two configurations of prepared states. One will serve as the target that is self-tested by a correspondingly defined witness. While we are interested in the symmetry of a configuration of 3 Bloch vectors, we need to add a fourth one here for the actual self-testing task, according to \eqref{eq:extraprep}. The first three Bloch vectors, at the same time, are parametrized as
\begin{align}\label{eq:target}
    \Vec{n_1} &= (0,0,1)\ &\ \Vec{n_2} &= (0, \sin{\theta_2},\cos{\theta_2}) \\ \nonumber
    \Vec{n_3} &= (\sin{\theta_3}\cos{\varphi_3}, \sin{\theta_3}\sin{\varphi_3},\cos{\theta_3})\ &\ \Vec{n_4} &= -\sum_{x=1}^3r_x\Vec{n_x}
\end{align}
where for the three parameters $\theta_2$, $\theta_3$ and $\varphi_3$ we make the unconventional choice $\theta_2 \in [0,\pi]$, $\theta_3 \in [0,2\pi)$ and $\varphi_3 \in [0,\pi)$ in order to represent $\varphi_3 = 90^{\circ}$ and what would be $\varphi_3 = 270^{\circ}$ by the same value. Irrespecitve of our unconventional choice, these parameters uniquely characterise any target.

On the other hand, we are also interested in symmetric configurations for which a given witness will be evaluated. According to what we said about the target, we need 4 Bloch vectors in total but only the first 3 of which are arranged in a symmetric way. Assuming that the symmetry plane is between $\Vec{n_2'}$ and $\Vec{n_3'}$, these four Bloch vectors are parametrized as
   \begin{align}\nonumber
    \Vec{n_1'} &= (0,0,1)\ &\ \Vec{n_2'} &= (\sin{\theta_2'}\cos{\phi_2'}, \sin{\theta_2'}\sin{\phi_2'},\cos{\theta_2'}) \\ \nonumber
    \Vec{n_3'} &= (-\sin{\theta_2'}\cos{\phi_2'}, \sin{\theta_2'}\sin{\phi_2'},\cos{\theta_2'})\ &\ \Vec{n_4'} &= (\sin{\theta_4'}\cos{\phi_4'}, \sin{\theta_4'}\sin{\phi_4'},\cos{\theta_4'}).
\end{align}
$\theta_2' \in [0,\pi],\ \phi_2' \in [0,2\pi],\ \theta_4' \in [0,\pi], \phi_4' \in [0,2\pi]$. Note that any permutation of the first three of these Bloch vectors will preserve the existence of a symmetry plane among them, so that all three permutations giving rise to different symmetry planes will be considered throughout our study. The fourth Bloch vector is just left arbitrary in any case as it does not play a role from our point of view of symmetry.

For self-testing the target according to the construction of Ref.~\cite{DrotosEtAl2024}, we initially choose $r_x=1$ for all $x$ for simplicity, and compute the witness coefficients as per the chosen construction, as well as the maximum witness value $Q_{max}$. Other choices of $r_x$ are also considered later.

We recall that certifying asymmetry of some unknown configuration requires exceeding the highest witness value $Q_{mirror}$ attainable by symmetric configurations. Therefore, following Ref.~\cite{TaoutiouiEtAl2025}, we determine $Q_{mirror}$ associated with each target and its corresponding witness functional, and also define the witness gap $\Delta = Q_{max}-Q_{mirror}$.

We will analyze the (three-dimensional) landscape of this witness gap as a function of the target configuration, as well as those of its consituents. We will pay special attention to differences arising from different choices about the self-testing construction.

For each point of the landscape, i.e., each target configuration, computing the witness matrix $w$ and the associated quantum maximum $Q_{max}$ is numerically straightforward and has been implemented with the GNU Scientific Library \cite{galassi2009gnu}. To numerically determine the corresponding value of $Q_{mirror}$, we have turned to the heuristic algorithm of Nelder and Mead \cite{NelderMead1965}, also using the GNU Scientific Library. Examples treated by SDP-based methods have confirmed that an upper bound is saturated by our heuristic solution with sufficient precision, see Appendix~\ref{sec:precision}.


\section{Findings}\label{sec:find}

\subsection{Analysis of a single witness construction choice}\label{sec:analysis_single}

In this subsection, we fix our choice, including all of the parameters, for constructing the witness; in particular, we choose $r_x = 1$ in the above-described construction for all $x \in \{1,2,3\}$ for simplicity. We study the gap as a function of the target configuration, which is charaterized by three parameters; for an easy visualization, we present landscapes of the gap over two parameters while keeping the third one fixed.

\begin{figure}[!h]
  \centering
  \includegraphics[width=0.7\linewidth]{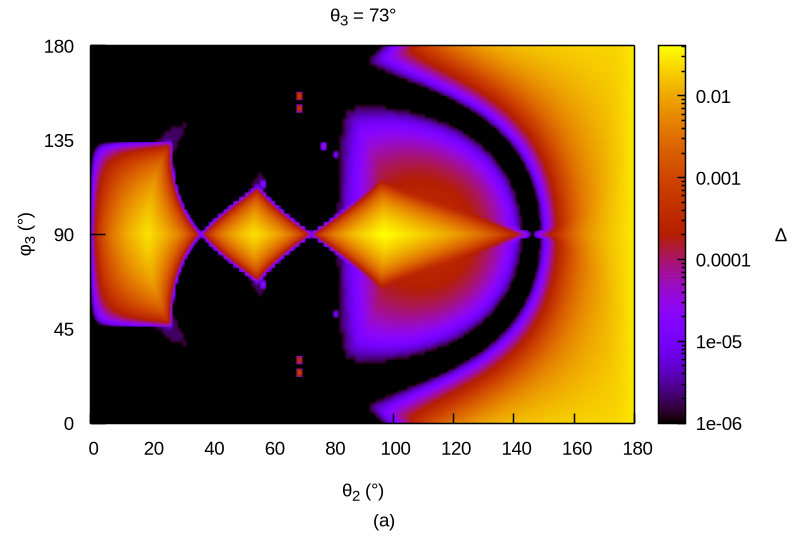}\\
  \includegraphics[width=0.7\linewidth]{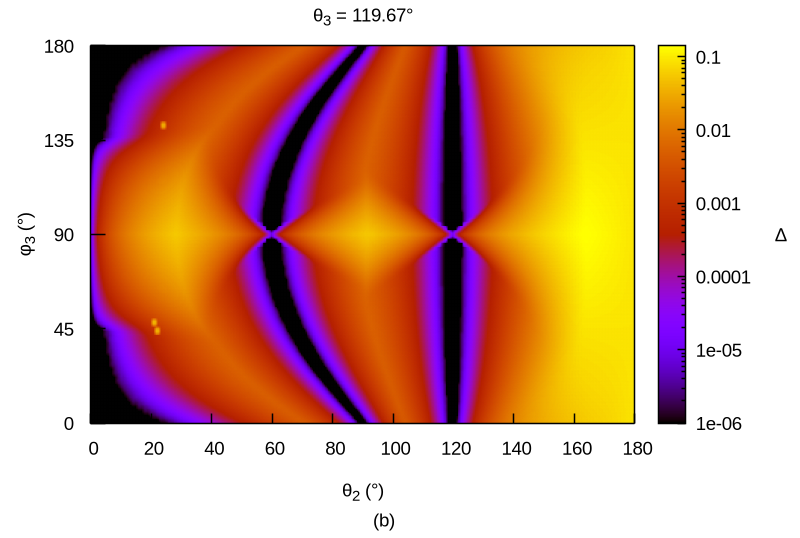}\\
  \includegraphics[width=0.7\linewidth]{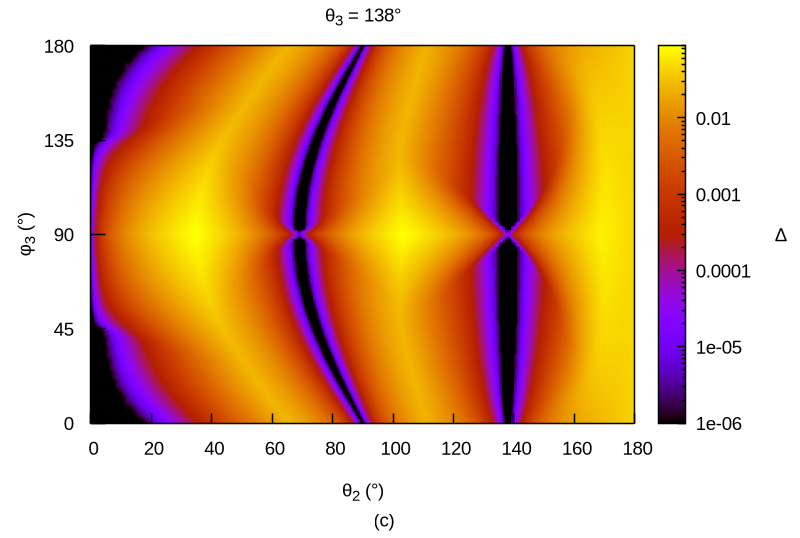}
  \caption{The gap $\Delta$, represented by the coloring on a logarithmic scale, as a function of the parameters $\theta_2$ and $\varphi_3$ of the target configuration for different fixed values of $\theta_3$ as indicated in the panels. $\theta_3 \approx 119.67^{\circ}$ corresponds to a global optimum. $r_x = 1$ for all $x \in \{1,2,3\}$.}
  \label{fig:t2p3delta}
\end{figure}

\begin{figure}[!h]
  \centering
  \includegraphics[width=0.7\linewidth]{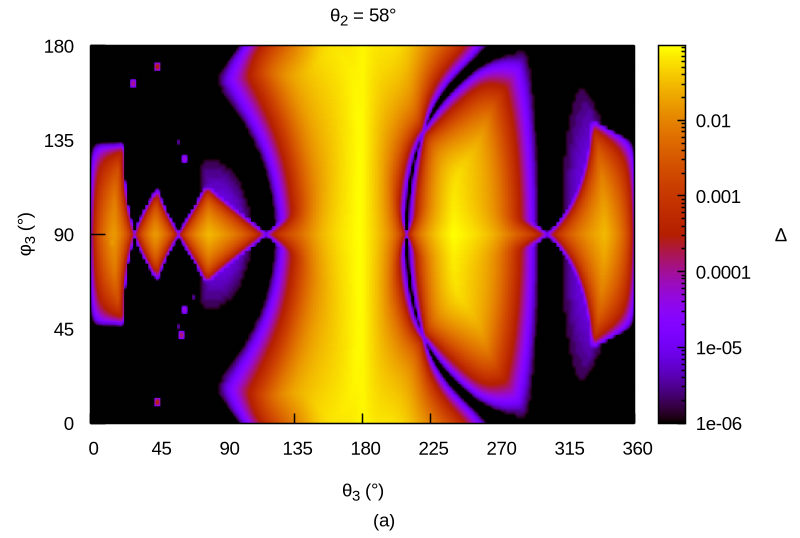}\\
  \includegraphics[width=0.7\linewidth]{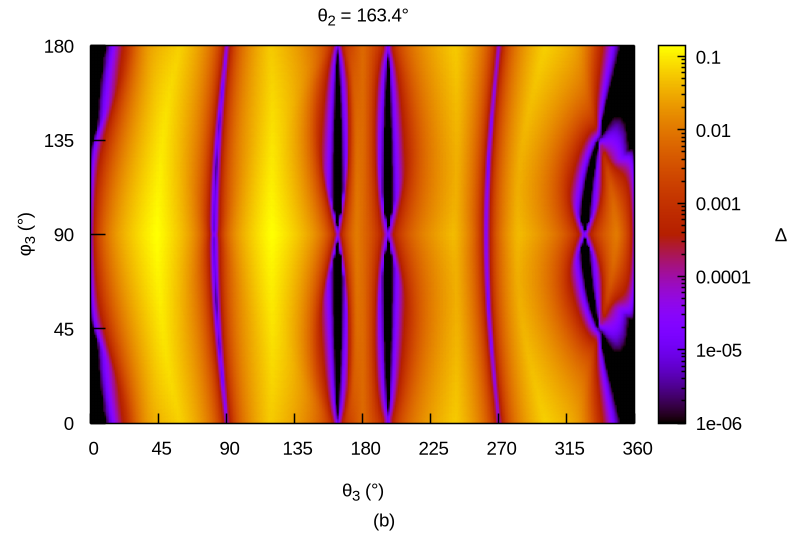}
  \caption{The gap $\Delta$, represented by the coloring on a logarithmic scale, as a function of the parameters $\theta_3$ and $\varphi_3$ of the target configuration for different fixed values of $\theta_2$ as indicated in the panels. $\theta_3 \approx 163^{\circ}$ corresponds to a global optimum. $r_x = 1$ for all $x \in \{1,2,3\}$.}
  \label{fig:t3p3delta}
\end{figure}

We begin with landscapes over $(\theta_2,\varphi_3)$ and $(\theta_3,\varphi_3)$ planes. Figures~\ref{fig:t2p3delta} and~\ref{fig:t3p3delta} illustrate that these landscapes numerically appear smooth in between curves where the gap is zero; these latter correspond to symmetric target configurations. Besides that, the only prominent characteristic is a maximum in $\theta_2$ or $\theta_3$ at constant $\varphi_3 = 90^{\circ}$, which appears to be independent of the value of the fixed parameter, and is approached arbitrarily by running another Nelder--Mead heuristic optimization, for the gap, in the full three-dimensional parameter space. While we haven't been able to analytically demonstrate either the smooth nature of the landscape or the $\varphi_3 = 90^{\circ}$ location of a ridge, we hypothesize them to be exact in order to proceed.

\begin{figure}[!h]
  \centering
  \includegraphics[width=0.7\linewidth]{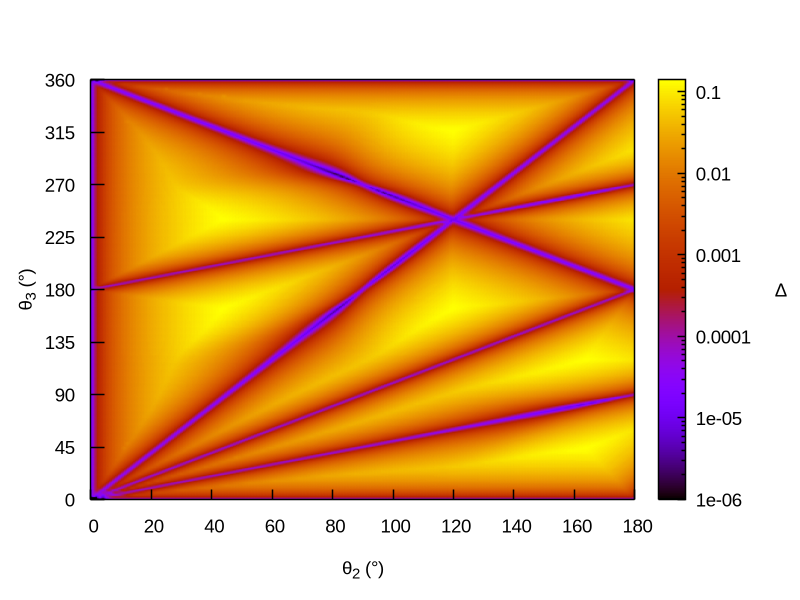}
  \caption{The gap $\Delta$, represented by the coloring on a logarithmic scale, as a function of the parameters $\theta_2$ and $\theta_3$ of the target configuration when $\varphi_3=90^{\circ}$ is fixed. $r_x = 1$ for all $x \in \{1,2,3\}$.}
  \label{fig:t2t3delta}
\end{figure}

Since we are mainly concerned about high values of the gap, from now on we will concentrate on the $(\theta_2,\theta_3)$ plane at $\varphi_3 = 90^{\circ}$. This landscape is numerically found to be smooth as well, except for lines of zero value, as illustrated by Fig.~\ref{fig:t2t3delta}. It is particularly easy in this case to read off the parameters of the mentioned lines of zero value (see Appendix~\ref{sec:zerolines}); then, a direct substitution into the parametrized form of the target configuration \eqref{eq:target} confirms the symmetric nature for these lines.

Notably, these lines divide the $(\theta_2,\theta_3)$ plane into six regions after taking into account periodic boundary conditions appropriately. (In particular, the lower half of the right-hand-side boundary, at $\theta_2 = 180^{\circ}$, is identified with its upper half.) We also point out that choosing $r_x = 1$ for all $x \in \{1,2,3\}$ results in the same set of witness coefficients $w_{xy}$ after any relabelling in $x$. As a consequence, the values $Q_{max}$ and $Q_{mirror}$ remain unchanged after such a relabelling. In total, there are $6$ possible labellings, so that almost every point in the landscape of the gap will have $5$ further equivalents.

The exception is when some given relabelling preserves the geometry of the configuration, which is so for \emph{symmetric} configurations and thus on the mentioned lines of zero value. Since any configuration can be transformed to any of its mirror images through continuous deformation, and the topology of lines of zero value can be described by closed loops only, we conclude that the six regions between the lines of zero value belong to the relabellings of the same geometry. Same geometry means in particular that the inner angles and the side lengths of the equivalent triangles are the same.

Fig.~\ref{fig:t2t3delta} suggests that there is exactly one maximum of the gap landscape in each of the $6$ regions. According to the above, these $6$ maxima are all global maxima and are equivalent in terms of relabelling. It is then sufficient to specify only one of them; in particular, assuming smoothness of the landscape, we find the parameters listed in Table~\ref{tab:opt} by means of the earlier-mentioned three-dimensional heuristic optimization in the parameter space. (Equivalence of these maxima in terms of relabelling has directly been checked as well.)

\begin{table}[ht]
      \centering
      \begin{tabular}{c|ccc|c}
          $r_1$ & $\theta_2$ & $\theta_3$ & $\varphi_3$ & $\Delta$ \\ \hline
          $1$ & $43.7^{\circ}$ & $163.4^{\circ}$ & $90^{\circ}$ & $0.143$\\
          $4$ & $39.4^{\circ}$ & $160.2^{\circ}$ & $90^{\circ}$ & $0.174$
      \end{tabular}
      \caption{Optimal target configurations with regard to the gap $\Delta$, in terms of parametrization~\eqref{eq:target}, for two different values $r_1$ of the witness construction. $r_2 = r_3 = 1$.}
      \label{tab:opt}
\end{table}

We already point out a relevant consequence here: the gap $\Delta$ must then be different for this optimal target configuration from that of any other configuration that defines a geometrically similar just, say, smaller triangle (which always exists). From a geometric point of view, it is not meaningful to say that one of two similar triangles is more asymmetric than the other; hence the gap $\Delta$, as in fact mentioned in Ref.~\cite{TaoutiouiEtAl2025}, cannot be regarded as a quantifier of symmetry only. We will see later that the dependence on size is not the only issue.

\begin{figure}[!h]
  \centering
  \includegraphics[width=0.7\linewidth]{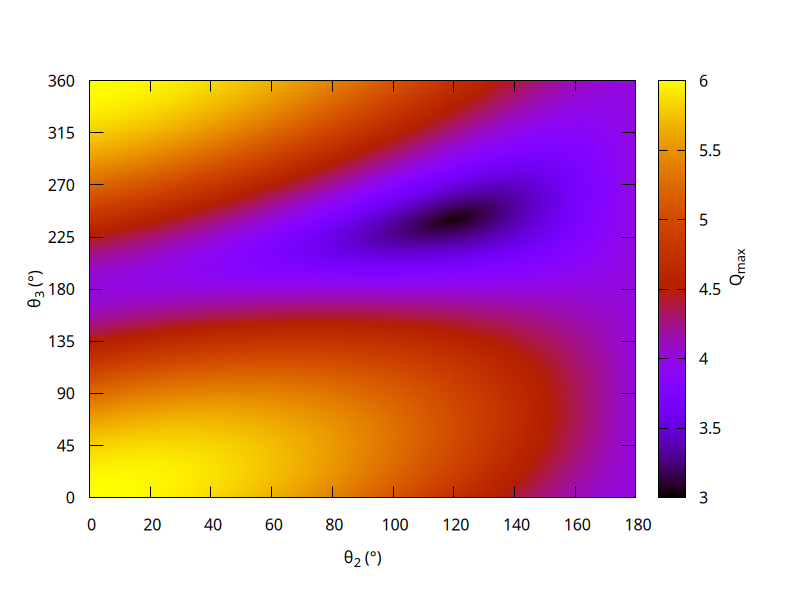}\\
  \includegraphics[width=0.7\linewidth]{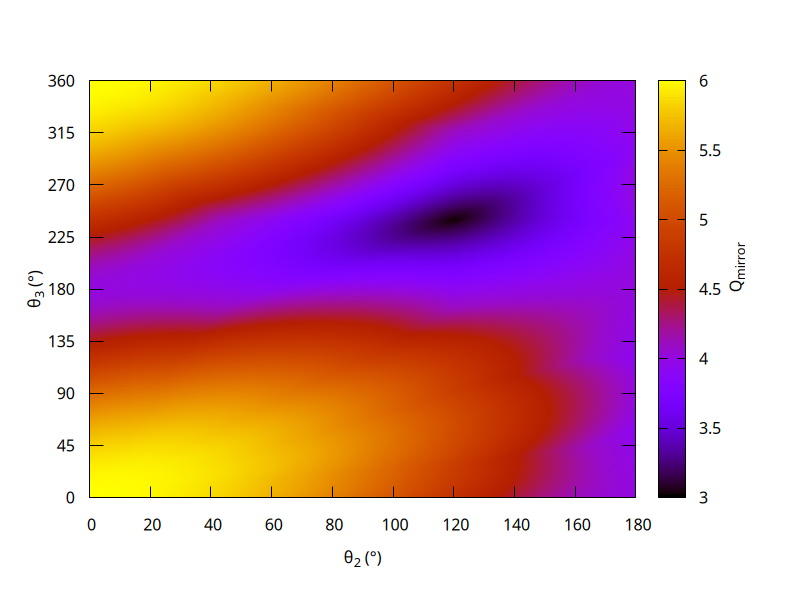}
  \caption{(a) The quantum maximum $Q_{max}$ and (b) the symmetry-constrained quantum maximum $Q_{mirror}$, represented by the coloring, as a function of the parameters $\theta_2$ and $\theta_3$ of the target configuration when $\varphi_3=90^{\circ}$ is fixed. $r_x = 1$ for all $x \in \{1,2,3\}$.}
  \label{fig:t2t3Q}
\end{figure}

For the moment, we will analyze further our example; in particular, what is special about the optimal target configuration. For this purpose, we plot the landscapes of $Q_{max}$ and $Q_{mirror}$ separately; remember that $\Delta$ is the difference of the former from the latter. According to Fig.~\ref{fig:t2t3Q}, we find that the values of both vary much more across the landscape than that of $\Delta$. These landscapes themselves appear to be smooth and to be mainly determined by a global-scale structure (which, although not obviously noticeable, is subject to the mentioned equivalence under relabelling). Any specialty about the six (equivalent) optima of $\Delta$, which lie at junctions of three ridges of the $\Delta$ landscape rising from junctions of the lines of zero values in Fig.~\ref{fig:t2t3delta}, only appears in the landscape of $Q_{mirror}$, in Fig.~\ref{fig:t2t3Q}b: the ridges of $\Delta$ just mentioned are accompanied by troughs of $Q_{mirror}$. In Fig.~\ref{fig:t2t3Q}a, showing the landscape of $Q_{max}$, no special feature can be seen at these locations.

This means that, from the point of view of our self-testing construction, these optimal target configurations are not special with respect to target configurations in their neighborhoods of any size. Instead, the witness function happens to be such that it poses some special limitation, not explained, near the relevant locations of the landscape on how closely the unconstrained optimum can be approached when allowing for mirror-symmetric configurations only. In this sense, we may say that these optima are more closely related to the symmetry-constrained configurations than to the actual target configurations self-tested by the chosen witness construction.

\subsection{An example for an altered parameter}\label{sec:altered}

Now let us see what is preserved from the above conclusions if we deviate from the $r_x = 1$, $x \in \{1,2,3\}$, choice. None of the target vectors is inherently distinguished by the construction, and it seems that altering a single $r_x$ provides sufficient information for our purposes, so that we simply take $r_1 = 4$ and $r_x = 1$ for $x \in \{2,3\}$.

\begin{figure}[!h]
  \centering
  \includegraphics[width=0.7\linewidth]{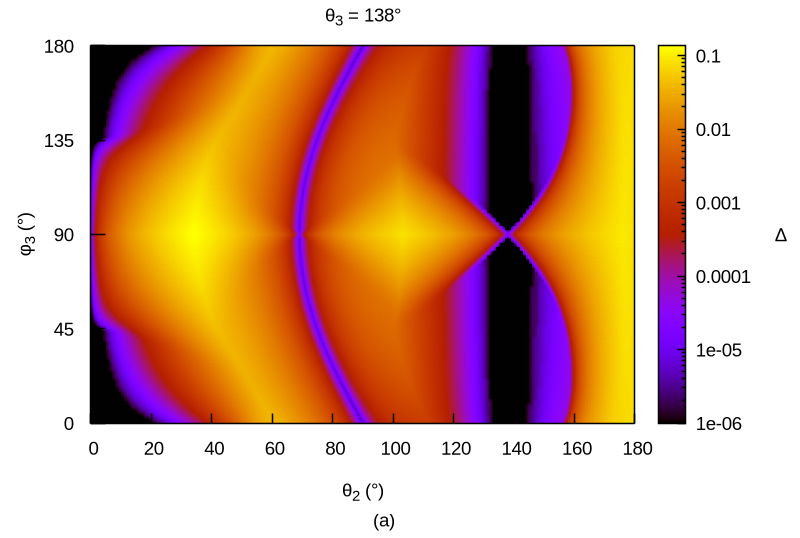}\\
  \includegraphics[width=0.7\linewidth]{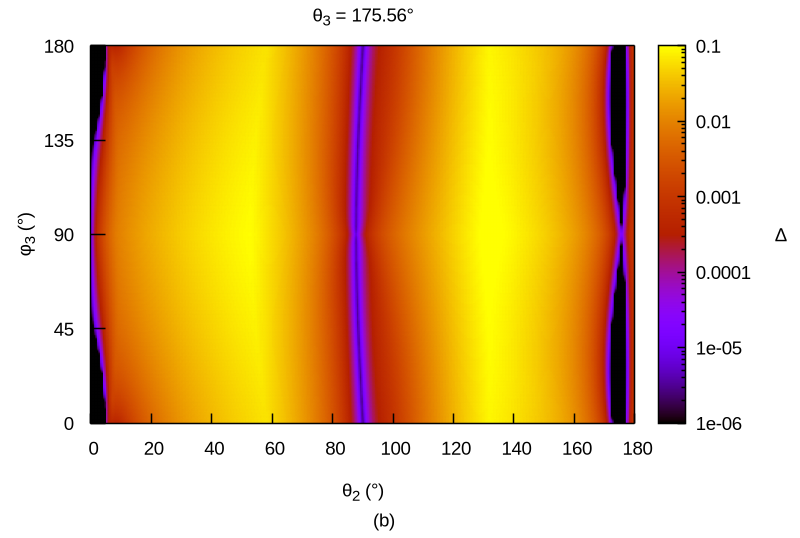}\\
  \caption{The gap $\Delta$, represented by the coloring on a logarithmic scale, as a function of the parameters $\theta_2$ and $\varphi_3$ of the target configuration for fixed values of $\theta_3$, as indicated in the panels. $\theta_3 \approx 175.56^{\circ}$ correspond to global optima. $r_1 = 4$ and $r_2 = r_3 = 1$.}
  \label{fig:t23p3delta_4}
\end{figure}

\begin{figure}[!h]
  \centering
  \includegraphics[width=0.7\linewidth]{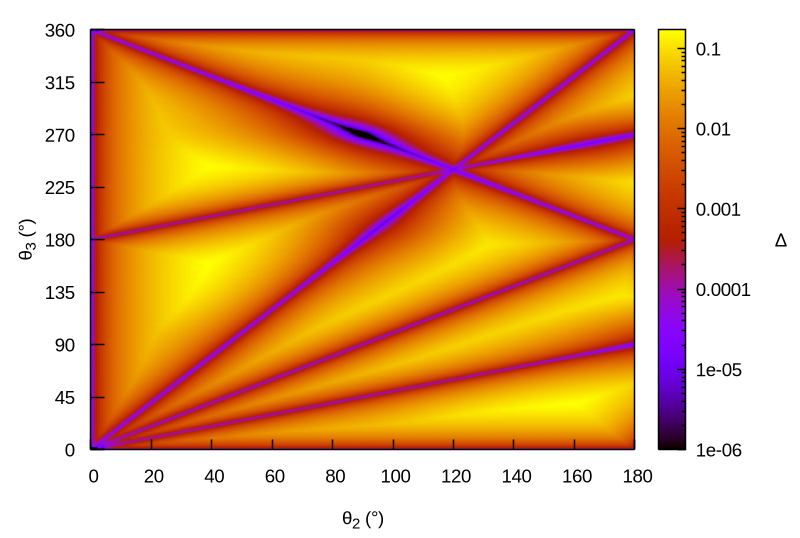}
  \caption{The gap $\Delta$, represented by the coloring on a logarithmic scale, as a function of the parameters $\theta_2$ and $\theta_3$ of the target configuration when $\varphi_3=90^{\circ}$ is fixed. $r_1 = 4$ and $r_2 = r_3 = 1$.}
  \label{fig:t2t3delta_4}
\end{figure}

Landscapes of the gap $\Delta$ over planes $(\theta_2,\varphi_3)$ and $(\theta_3,\varphi_3)$ indicate that $\varphi_3 = 90^{\circ}$ remains optimal in the sense discussed in the previous section; see Fig.~\ref{fig:t23p3delta_4} for some examples. The landscape of $\Delta$ over the $(\theta_2,\theta_3)$ plane at $\varphi_3 = 90^{\circ}$ remains very similar to that observed for $r_1 = 1$, one may compare Figs.~\ref{fig:t2t3delta_4} and~\ref{fig:t2t3delta} to see this.

Now we do not have a reason to think that the six regions of the landscape separated by the lines of zero value would be equivalent. Since $r_1 \neq r_2 = r_3$, every labelling that associates a different vector with index $1$ in the same geometric configuration of the three vectors is expected to result in a different value of $Q_{max}$. Hence, the six regions should belong to three different equivalence classes, which is confirmed by our numerics.

One global optimum is found at a different location from the corresponding location for $r_1 = 1$, see Table~\ref{tab:opt}. While the difference in the location is not big, our numerics is sufficiently precise (due to confirmation by methods based on semi-definite programming (SDP); see Appendix~\ref{sec:precision}) to provide a mathematically precise proof for its existence.

This finding implies that the geometry of the target configuration associated with the largest gap depends on a parameter choice in the self-testing construction. As a consequence, the gap may not be a characteristic of the geometry of the configuration.

\subsection{Adaptive parameter choice for a normalized landscape}

While it seems that we need to abandon the idea of an association between the gap and the geometry of the target configuration, the gap is still expected to be useful to quantify how difficult it is, in experiments or applications, to certify a violation of the relevant constraint, which is mirror symmetry in our case. However, we have seen in Fig.~\ref{fig:t2t3Q} that the relevant values $Q_{max}$ and $Q_{mirror}$ of the witness may vary considerably over the landscape (either the full three- or already any two-dimensional one), which will thus affect the interpretation of a ``large'' or a ``small'' gap.

Difficulty in certification in real-world implementations arises, apart from finite sample sizes for stastistics, from different kinds of noise which will decrease the evaluated value of the witness, possibly below the constrained (in our case, mirror-symmetric) bound. Arguably, noise (affecting on the unknown configuration to be certified) will have a larger numerical effect if the target witness value is higher itself. Therefore, normalization of the target witness value, $Q_{max}$, appears to be a reasonable proxy for keeping noise constant over the parameter space corresponding to a given witness construction scheme.

In our case, keeping $Q_{max}$ normalized over the landscapes studied so far is easy by an appropriate rescaling of the parameters $r_x$, all of them with the same factor, $Q_{max}$, simultaneously. By this, we ensure a new $\tilde{Q}_{max} = 1$ over the entire landscape, and $Q_{mirror}$ is replaced by its original value divided by the old $Q_{max}$; the new gap $\tilde{\Delta}$ will be obtained from the old witness values $Q_{max}$ and $Q_{mirror}$ as $\tilde{\Delta} = 1-Q_{mirror}/Q_{max}$.

\begin{figure}[!h]
  \centering
  \includegraphics[width=0.7\linewidth]{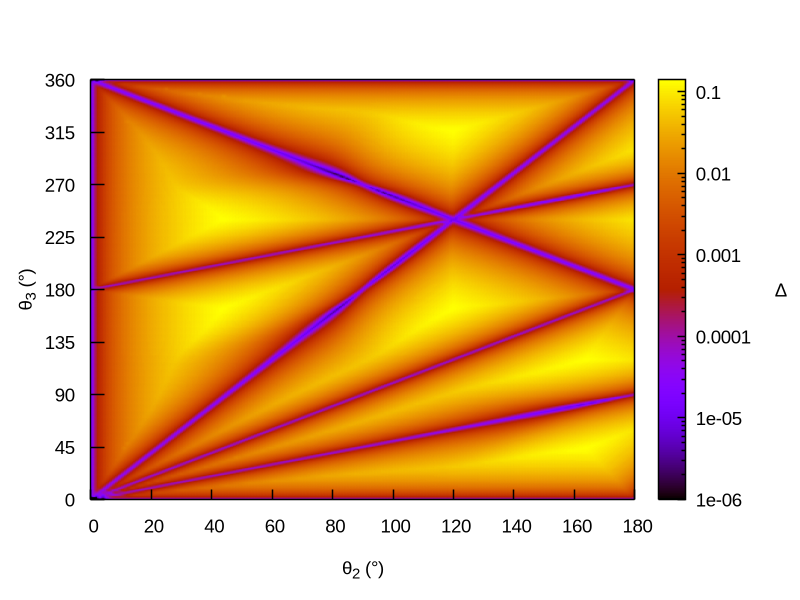}
  \caption{Same as Fig.~\ref{fig:t2t3delta} for $\tilde{\Delta}$ instead of $\Delta$.}
  \label{fig:t2t3delta_new}
\end{figure}

The landscape of $\tilde{\Delta}$ associated with Fig.~\ref{fig:t2t3delta} is shown in Fig.~\ref{fig:t2t3delta_new}. In fact, besides that we find no qualitative difference, the locations of the relevant maxima are not distinguishable at our numerical precision from those of their old counterparts. This is surprising from the point of view that, from elementary algebra, some difference in the location is generically expected whenever the landscape of $\Delta$ is differentiable at a given maximum. While further exploration of the issue would be justified in order to assess the relevance of normalization in a generic construction, analyzing the present construction for self-testing has mostly been exhausted from the paper's point of view, and we will now turn to discussing our findings and drawing conclusions.

\section{Discussion, conclusion}\label{sec:disc_concl}

In the last section, we analysed properties of the landscape of the witness gap over the (three-dimensional) parameter space of the target configuration, then we concluded that quantitative details depend on choices such as a simple parameter choice of the underlying self-testing construction. To place this finding in a wider context, we recall the target configuration with the largest gap from Ref.~\cite{TaoutiouiEtAl2025} which uses a different construction for self-testing. We compare that target configuration with one of ours corresponding to the largest gap in Table~\ref{tab:comp_Malek} and conclude that they are substantially different. Our construction does not even preserve the special property of collinearity between two of the target vectors.

\begin{table}[ht]
      \centering
      \begin{tabular}{ccc|c}
          $\alpha_{12}$ & $\alpha_{13}$  & $\alpha_{23}$ \\ \hline
          $43.7^{\circ}$ & $ 119.7^{\circ}$  & $196.6^{\circ}$ \\
          $58.4^{\circ}$ & $ 121.6^{\circ}$  & $180.0^{\circ}$ \\
      \end{tabular}
      \caption{Angles $\alpha_{ij}$ between target Bloch vectors $\Vec{n_i}$ and $\Vec{n_j}$ in the target configuration corresponding to the largest gap in the self-testing construction of the present article (top row) and that of Ref.~\cite{TaoutiouiEtAl2025} (bottom row).}
      \label{tab:comp_Malek}
\end{table}

This experience reinforces our conclusion that the gap cannot be used as a quantifier of asymmetry alone. In general, we can say that its value is dependent on the definition of the witness, and, as discussed in Section~\ref{sec:analysis_single}, also on the size of otherwise similar triangles defined by the target Bloch vectors.

On the other hand, we have made the important numerical observation that the landscape of the witness gap, irrespective of the parameter choice, is smooth and has a simple shape. Comparing different parameters also suggests that it is structurally robust with respect to alterations in parameters. These properties make it possibly useful for practical purposes.

Regarding ``practical applications'', whatever the actual goal of certifying asymmetry or some other qualitative property of an unknown configuration is, a larger gap for the target is intuitively expected to make certification easier, which is then relevant for appropriate choices of a target and parameters. We will explore tolerance against imperfections in future work.

\section*{Acknowledgements}

Useful comments about the manuscript by A. Taoutioui and T. Vértesi are gratefully acknowledged. We acknowledge the support of the EU (CHIST-ERA MoDIC) along with the National Research, Development and Innovation Office NKFIH (No. \verb+2023-1.2.1-ERA_NET-2023-00009+). G.D. acknowledges support from the European Union (European Social Fund Plus) and the Government of the Balearic Islands through a Vicenç Mut postdoctoral fellowship (No. PD-035-2023).

\section*{Code and data availability}

The code and data that support the findings of this study are openly available under Ref.~\cite{accdata}.

\appendix
\section{Lines of zero gap value}\label{sec:zerolines}

It is easy to analytically prove the existence of lines with $\Delta=0$ in Fig.~\ref{fig:t2t3delta} (or, in fact, in the $\Delta$ landscape over the $(\theta_2,\theta_3)$ plane at $\varphi_3 = 90^{\circ}$ for any possible construction of the witness). $\Delta=0$ is equivalent to saying that the target configuration is symmetric for the given parameters, which we can actually check.

In particular, we can read off, from Fig.~\ref{fig:t2t3delta}, parameters of lines in the plane $(\theta_2,\theta_3)$ where we conjecture symmetric target configurations. We parametrize the lines as
    \begin{equation}\label{eq:zerolines}
        \theta_3 = m\theta_2 + b ;
    \end{equation}
then, if we substitute $\theta_3$ according to \eqref{eq:zerolines} into \eqref{eq:target} (which defines the parametrization of the target Bloch vectors), we can algebraically confirm the symmetry of the target configuration with the line parameters listed in Table~\ref{tab:zerolines}.

\begin{table}[ht]
    \centering
    \begin{tabular}{c|c} 
        $a$ & $b$\\\hline
        $\frac{1}{2}$ & $0$\\
        $1$ & $0$\\
        $2$ & $0$ \\
        $\frac{1}{2}$ & $\pi$\\
        $-1$ & $2\pi$\\
        $a \xrightarrow{}+\infty$ & $0$ \\
    \end{tabular}
    \caption{The $a$ and $b$ parameters of \eqref{eq:zerolines} for the lines of zero gap value.}
    \label{tab:zerolines}
\end{table}

\section{Precision of our numerics}\label{sec:precision}

Since we are relying heavily on numerics in our paper, including a few main claims, we need to assess its precision.

Target configurations enter into the computations as inputs, and the corresponding quantum maxima $Q_{max}$ are essentially obtained in closed form, so that machine precision can be assumed here. $Q_{mirror}$, however, is subject to uncertainty because of using a heuristic optimization algorithm, that of Nelder and Mead \cite{NelderMead1965}, for finding it.

Similarly as in Ref.~\cite{TaoutiouiEtAl2025}, one could use an algorithm based on semi-definite programming (SDP) for finding the optimum; however, it is computationally too expensive for our purposes. Instead, we run SDP for a few selected target configurations only, to show that the heuristic results are sufficiently reliable. In particular, we apply order 3 in Lasserre's relaxation scheme \cite{Lasserre2001} in the YALMIP package for MATLAB \cite{YALMIP}, calling SDPT3 \cite{TohEtAl1999,TutuncuEtAl2003} as the SDP solver.

\begin{figure}[!h]
  \centering
  \includegraphics[width=0.7\linewidth]{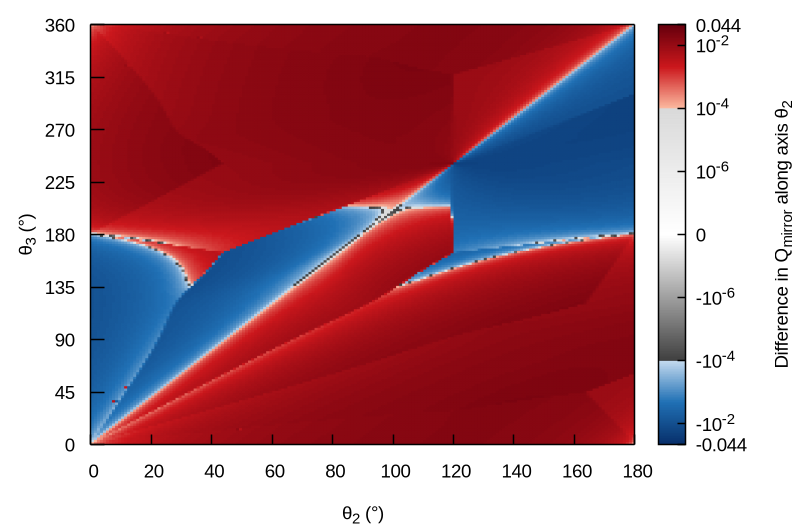}\\
  \includegraphics[width=0.7\linewidth]{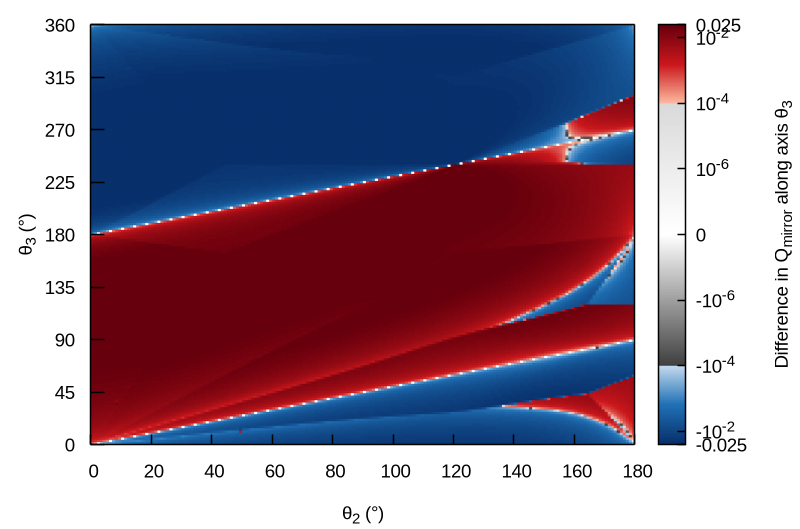}\\
  \caption{Differences in $Q_{mirror}$, as a function of $\theta_2$ and $\theta_3$ of the target configuration when $\varphi_3=90^{\circ}$, between neighboring grid points along axis (a) $\theta_2$ and (b) $\theta_3$ represented by the coloring on a logarithmic scale. Grayscale is applied for small magnitudes which appear to conveniently cover numerical precision, cf. Table~\ref{tab:num_r1_1}. $r_1=r_2=r_3=1$.}
  \label{fig:t2t3diff}
\end{figure}

We start with a number of randomly selected locations from the landscape of $Q_{mirror}$ underlying Fig.~\ref{fig:t2t3delta}. The examples presented in Table~\ref{tab:num_r1_1} confirm that precision is much higher than differences in values of $Q_{mirror}$ between typical neighboring grid points, as presented in Fig.~\ref{fig:t2t3diff}, so that the smoothness and the shape of the landscape at the given resolution are not artefacts. (Rarely, we can identify cases when the SDP upper bound on $Q_{mirror}$ is lower than the Nelder--Mead value, but the difference is within numerical precision as set for SDPT3.) To see that this is independent of the special parameter choice $r_i=1$ for $i\in\{1,2,3\}$, we make a similar comparison in Table~\ref{tab:num_r1_4} for the same locations corresponding to the $Q_{mirror}$ landscape underlying Fig.~\ref{fig:t2t3delta_4}.

\begin{table}[ht]
      \centering
      \begin{tabular}{ccc|ccc}
          $\theta_2$ & $\theta_3$ & $\varphi_3$ & $Q_{mirror}$ (NM LB) & $Q_{mirror}$ (SDP UB) \\ \hline
$50^{\circ}$ & $73^{\circ}$ & $85^{\circ}$ & $5.58144701$ & $5.58144701$ \\
$120^{\circ}$ & $73^{\circ}$ & $70^{\circ}$ & $4.96164898$ & $4.96164899$ \\
$58^{\circ}$ & $135^{\circ}$ & $70^{\circ}$ & $4.73796771$ & $4.73796773$ \\
$10^{\circ}$ & $45^{\circ}$ & $90^{\circ}$ & $5.82477291$ & $5.82477293$ \\
$10^{\circ}$ & $135^{\circ}$ & $90^{\circ}$ & $4.54424525$ & $4.54424526$ \\
$20^{\circ}$ & $90^{\circ}$ & $90^{\circ}$ & $5.32935193$ & $5.32935196$ \\
$20^{\circ}$ & $150^{\circ}$ & $90^{\circ}$ & $4.33408461$ & $4.33408463$ \\
$40^{\circ}$ & $135^{\circ}$ & $90^{\circ}$ & $4.66571088$ & $4.66571091$ \\
$80^{\circ}$ & $180^{\circ}$ & $90^{\circ}$ & $3.98985610$ & $3.98985613$ \\
      \end{tabular}
      \caption{Different numerical approaches to the maximum $Q_{mirror}$: heuristic Nelder--Mead optimization (NM LB, as this is a lower bound on the exact value) and upper bound from SDP (SDP UB), see text. The control parameters are defined in Eq.~\eqref{eq:target}. $r_i = 1$ for $i \in \{1,2,3\}$.}
      \label{tab:num_r1_1}
\end{table}

\begin{table}[ht!]
      \centering
      \begin{tabular}{ccc|ccc}
          $\theta_2$ & $\theta_3$ & $\varphi_3$ & $Q_{mirror}$ (NM LB) & $Q_{mirror}$ (SDP UB) \\ \hline
$50^{\circ}$ & $73^{\circ}$ & $85^{\circ}$ & $11.21486253$ & $11.21486253$ \\
$120^{\circ}$ & $73^{\circ}$ & $70^{\circ}$ & $10.19190668$ & $10.19190668$ \\
$58^{\circ}$ & $135^{\circ}$ & $70^{\circ}$ & $10.11338453$ & $10.11338454$ \\
$10^{\circ}$ & $45^{\circ}$ & $90^{\circ}$ & $11.74749833$ & $11.74749833$ \\
$10^{\circ}$ & $135^{\circ}$ & $90^{\circ}$ & $10.35526277$ & $10.04004902$ \\
$20^{\circ}$ & $90^{\circ}$ & $90^{\circ}$ & $11.07202183$ & $11.07202184$ \\
$20^{\circ}$ & $150^{\circ}$ & $90^{\circ}$ & $10.11143541$ & $10.11143543$ \\
$40^{\circ}$ & $135^{\circ}$ & $90^{\circ}$ & $10.19060861$ & $10.19060861$ \\
$80^{\circ}$ & $180^{\circ}$ & $90^{\circ}$ & $9.31918333$ & $9.31918334$ \\
      \end{tabular}
      \caption{Same as Table~\ref{tab:num_r1_1} but for $r_1 = 4$ and $r_2 = r_3 = 1$}
      \label{tab:num_r1_4}
\end{table}

Most interesting is to see whether the difference between the locations of the optima presented in Table~\ref{tab:opt} may be an artefact. Evaluating $Q_{mirror}$ for both parameter choices at both locations in Table~\ref{tab:num_comp} shows that it is not so, which confirms this difference as a main claim supporting our view presented in Sections~\ref{sec:altered} and~\ref{sec:disc_concl}.

\begin{table}[ht!]
      \centering
      \begin{tabular}{c|ccc|ccc}
          $r_1$ & $\theta_2$ & $\theta_3$ & $\varphi_3$ & $Q_{mirror}$ (NM LB) & $Q_{mirror}$ (SDP UB) & $Q_{max}$ \\ \hline
          $1$ & $43.7^{\circ}$ & $163.4^{\circ}$ & $90^{\circ}$ & $4.09765782$ & $4.09765784$ & $4.240$ \\
          $4$ & $43.7^{\circ}$ & $163.4^{\circ}$ & $90^{\circ}$ & $9.74491552$ & $9.74491553$ & $4.280$ \\
          $1$ & $39.4^{\circ}$ & $160.2^{\circ}$ & $90^{\circ}$ & $4.16349734$ & $4.16349736$ & $9.889$ \\
          $4$ & $39.4^{\circ}$ & $160.2^{\circ}$ & $90^{\circ}$ & $9.78026457$ & $9.78026458$ & $9.954$ \\
      \end{tabular}
      \caption{Comparison of the maximum $Q_{mirror}$ in terms of different numerical approaches, at optimal locations for $r_1 = 1$ and $r_1 = 4$ along with $r_2 = r_3 = 1$. $Q_{max}$, entering the gap $\Delta$, is shown for reference.}
      \label{tab:num_comp}
\end{table}

\newpage
\printbibliography

\end{document}